\documentclass[10pt,a4paper]{article}
\usepackage[cp1251]{inputenc}
\usepackage[russian]{babel}
\usepackage[T2A]{fontenc}
\begin{document}
\vspace*{2cm}
\begin{center}{\Large \bf Geometrization of Classical Wave Fields}\end{center}
\par
\begin{center}{\large Oleg A.Olkhov} \end{center}

\begin{center} {\it N.N.Semenov Institute of Chemical Physics, Kosygin str.,4,
119991 Moscow, Russia; olega@gagarinclub.ru}\end{center}
\par\medskip
\begin{center}{AIP Conf.Proc.V.962. p.316, QUANTUM THEORY: Reconsideration of Foundations-4, Vaxjo, Sweden, 11-16 June 2007 (http://proceedings.aip.org/proceedings)}\end{center}
\par\medskip
\begin{center}\parbox{159mm}{{\small \noindent {\bf Abstract} New geometrical
interpretation of quantum physics is suggested. It is shown that the
Dirac equation for free quantum particles can be considered as a
relation describing propagation of the space topological defect with
wave and corpuscular properties. Such interpretation explains all
properties of quantum formalism that seems "strange" within
classical representations: appearance of probabilities, absence of
the particles trajectories, unremovable influence of the measurement
procedure, instantaneous nonlocal correlation in EPR--experiments.
The same interpretation is suggested for the Maxwell equations for
free electromagnetic waves. This interpretation explains
independence of light velocity on the source movement.}

}\end{center}
\par\medskip
\begin{center} {\bf INTRODUCTION}\end{center}

 We will show within unified approach, that the free matter field
(quantum particle with spin $1/2$) and free electromagnetic waves
can both be considered not as something existing into the space that
plays the role of scene but as a propagating topological defects of
the space itself. These defects appear to be embedded into
five--dimensional space, and observable quantum object and
electromagnetic wave are intersections of the defects with the
three-dimensional physical space. Their wave properties are a
consequence of the periodical movement of the defect in the "outer"
space--just this movement attributes phase to the particle and so
defines its wave properties. The classical notions of mass, energy
and momentum are expressed through the parameters with
dimensionality of length, and the Planck constant and the light
velocity play the role of coefficients of transfer from one system
of units to the other. It is shown that light velocity can be
considered as a topological invariant, and this is the reason for
its independence of the source motion. The uncertainty of shapes of
topological manifolds is the reason of the appearance of
probabilistic description.

We start with the geometrization of matter field and, doing so, we
have to keep in mind the exceptional accuracy of the modern quantum
formalism. Therefore, we suppose that attempts to find out new
geometrical description of quantum objects have to begin not with
the creation of a new formalism but with finding out a geometrical
interpretation of the well-known quantum equations whose validity is
beyond question. This is the reason why we start with an attempt to
find out geometrical interpretation of the Dirac equation for free
particle. Preliminary results were published earlier [1].
\par\medskip
\begin{center}{\bf TOPOLOGICAL INTERPRETATION OF THE DIRAC EQUATION}\end{center}
\par\medskip

This equation has the following symbolic form (see, e.g.,[2])
$$i\gamma^{\mu}\partial_{\mu}\psi=m\psi,\eqno (1)$$
where $\partial_{\mu}=\partial/\partial x_{\mu},\quad \mu
=1,2,3,4$,\quad $\psi (x)$ is the four-component Dirac bispinor,
$x_{1}=t, x_{2}=x, x_{3}=y$, $x_{4}=z$, and $\gamma^{\mu}$ are
four-row Dirac matrices. The summation in Eq.(1) goes over the
repeating indices with a signature $(1,-1,-1,-1)$. Here, $\hbar
=c=1$. For definite values of 4-momentum $p_{\mu}$, the solution to
Eq.(1) has the form of the plane wave
$$\psi=u(p_{\mu})\exp (-ip_{\mu}x^{\mu}),\eqno(2)$$ where $u(p_{\mu})$ is a normalized bispinor.
Substitution of (2) in Eq.(1) gives:
$$p_{1}^{2}-p_{2}^{2}-p_{3}^{2}-p_{4}^{2}=m^{2}.\eqno(3)$$

Different wave fields are described by different tensors: they
realize different representations of the Lorentz group (group of
4-rotation). Dirac bispinor consists of two--component spinors that
realizes two--dimensional representation of the Lorentz group, and
this is the formal reason to classify the Dirac wave field of Eq.(1)
as the object with "internal" angular momentum whose Z--projection
equals to $\pm 1/2$ [3]. Let us, firstly, consider transformation
properties of the Dirac bispinor from the other point of view. It is
known that there may be established correspondence between every
kind of tensors and some class of geometrical objects in the sense
that these tensors define invariant properties of above objects. For
example, usual vectors correspond to simplest geometrical
objects--to points [4], and this is the reason why Newtonian
mechanics uses vectors within its formalism. From this point of view
spinors correspond to nonorientable geometrical object (see, e.g.,
[5]). Therefore, we suppose that spinors are used in Eg.(1), because
this equation describes some nonorientable geometrical object and
"$spin = 1/2$" is a formal expression of the nonorientable property
of the object.

The above assumption have to be considered as the starting
hypothesis only. To define properties of the proposed geometrical
object more exactly we consider more precisely the symmetry
properties of the solution of Eq.(1). We rewrite function (2) and
relation (3) in the form
$$\psi=u(p_{\mu})\exp (-2\pi ix^{\mu}\lambda_{\mu}^{-1}).\eqno(4)$$
$$\lambda_{1}^{-2}-\lambda_{2}^{-2}-\lambda_{3}^{-2}-\lambda_{4}^{-2}=\lambda_{m}^{-2},\quad \lambda_{\mu}=2\pi p_{\mu}^{-1},
 \quad \lambda_{m}=2\pi m^{-1}.\eqno(5)$$
Let us now consider $\lambda_{\mu}$ as some parameters with a
dimensionality of length that have nothing to do with any wave
process into the space. Function (4) is an invariant with respect to
coordinates transformations
$$x^{'}_{\mu}=x_{\mu}+n_{\mu}\lambda_{\mu}, \quad n_{\mu}=0,\pm 1, \pm 2, ...\quad .\eqno (6)$$
Transformations (6) can be considered as elements of the discrete
group of translations operating in the 4-space where wave function
(4) is defined. Then function (4) can be considered as a vector
realizing this group representation.

As a bispinor, function (4) realizes representation of one more
group of the symmetry transformation of 4-space. Being a
four-component spinor, $\psi(x)$ is related to the matrices
$\gamma^{\mu}$ by the equations (see, e.g. [3])
$$\psi^{'}(x^{'})=\gamma^{\mu}\psi (x),$$ where $x\equiv
(x_1,x_2,x_3,x_4)$, and $x^{'}\equiv (x_1,-x_2,-x_3,-x_4)$ for $\mu
=1,x^{'}\equiv (-x_1,x_2,-x_3,-x_4)$ for $\mu =2$, and so on. This
means that the matrices  $\gamma^{\mu}$ are the matrix
representation of the group of reflections along three axes
perpendicular to the $x_\mu$ axis, and the Dirac bispinors realize
this representation [6].

Taken together, above two groups form a group of four sliding
symmetries with perpendicular axes (sliding symmetry means
translations plus corresponding reflections [7]). The physical
space-time does not have such symmetry.So, this group may operate
only in some auxiliary space. From the other hand, it is known that
discrete groups operating in some space can reflect a symmetry of
geometrical objects that have nothing in common with this space. It
will be the case when such space is a universal covering space of
some closed topological manifold. Universal covering spaces are
auxiliary spaces that are used in topology for the description of
closed manifolds, because discrete groups operating in these spaces
are isomorphic to fundamental groups of manifolds---groups whose
elements are different classes of closed pathes on manifolds (so
called $\pi_{1}$ group [8,9]). And we assume now, that function (4)
realizes a representation of the fundamental group of some closed
nonorientable topological 4-manifold---a specific curved part of the
space-time. Eq.(1) describes this manifold and imposes limitations
(5) on the possible values of the fundamental group parameters
$\lambda_{\mu}$. Space-time plays also the role of an universal
covering space for above manifold.

How above manifold can describe something moving in the space ?. At
the present time, only two-dimensional euclidean closed manifolds
are classified in details, and their fundamental groups and
universal covering planes are identified [8]. Therefore, we have no
opportunity for rigorous consideration of specific properties of
suggested psudoeuclidean 4-manifold. But qualitative properties,
explaining main ideas of new interpretation, can be investigated
using one of the advantages of geometrical approach---possibility of
employment of low-dimensional analogies. Using these analogies we
will show within elementary topology that above mentioned 4-manifold
represents propagation of the space topological defect that can
demonstrate specific properties of quantum particle represented by
solution (2): stochastic behavior and wave-corpuscular dualism.
\par\medskip
\begin{center} {\bf STOCHASTIC BEHAVIOR}\end{center}
\par\medskip

Consider the simplest example of closed topological
manifolds---one-dimensional manifold homeomorphic to a circle with
given perimeter length $\lambda$. The closed topological manifold is
representable by any of its possible deformations (without pasting)
that conserve manifold's continuity, and we will see that just this
property explains appearance of probabilities in quantum formalism.
For simplicity we consider plane deformations of the circle(some of
the possible deformations are shown at Fig.1).
\begin{center}
\begin{picture}(110,50)
\put(0,25){\circle{25}} \put(35,25){\oval(15,30)}
\put(75,25){\oval(35,10)}
\qbezier(110,35)(165,25)(110,20)\qbezier(110,35)(190,25)(110,20)\qbezier(165,25)(170,40)(175,30)
\qbezier(165,25)(165,30)(170,15)\qbezier(175,30)(180,35)(190,25)\qbezier(170,15)(175,27)(190,25)
\end{picture}
\end{center}

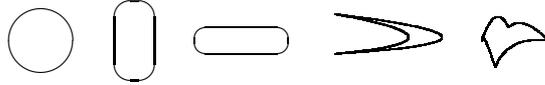
\noindent {\bf FIGURE 1.} Stochastic behavior.
\par\medskip

To use concrete simple calculations, we consider only all possible
manifold's deformations that have a shape of ellipse with perimeter
length $\lambda$. The equation for the ellipse on an euclidean plane
has the form
$$x^2/a^2+y^2/b^2=1, \eqno (7)$$where all possible values of the
semiaxes $a$ and $b$ are connected with the perimeter length
$\lambda$ by the known approximate relation
$$\lambda\simeq \pi[1,5(a+b)-(ab)^{1/2}].\eqno (8)$$This means that the
range of all possible values of $a$ is defined by the inequality
$a_{min}\leq a\leq a_{max}\simeq \lambda/1,5\pi, \quad a_{min}\ll
a_{max}$.

In the pseudoeuclidean two-dimensional "space-time," the equation
for our ellipses has the form (after substitution $y=it$)
$$x^2/a^2-t^2/b^2=1, \eqno (9)$$ and this equation defines the
dependence on time $t$ for a position of the point $x$ of the
manifold corresponding to definite $a$. At $t=0,x=\pm a$; that is,
our manifold is represented by the two point sets in one-dimensional
euclidean space, and the dimensions of these point sets are defined
by all possible values of $a$. So, at $t=0$, the manifold is
represented by two regions of the one--dimensional euclidean space
$a_{min}\leq| x|=a\leq a_{max}$. It can easily be shown that at
$t\neq 0$ these regions increase and move along the x-axis in
opposite directions \newpage(at Fig.2 this movement is shown only for
positive direction).
\begin{center}
\begin{picture}(250,40)
\put(0,25){\vector(1,0){230}}\thicklines{
\put(5,25){\line(1,0){30}}}\thicklines{\put(55,25){\line(1,0){50}}}
\thicklines{\put(125,25){\line(1,0){70}}}\thinlines{\put(35,35){\vector(1,0){18}}
\put(105,35){\vector(1,0){18}}\put(195,35){\vector(1,0){18}}\put(235,25){x}}
\put(20,35){$t_0$}\put(80,35){$t_1$}\put(160,35){$t_2$}
\end{picture}
\end{center}
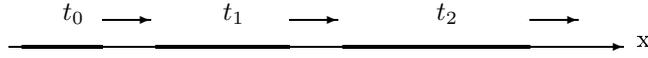
\noindent {\bf FIGURE 2.} Propagation of the spreading region of the
space.
\par\medskip

All another possible deformations of our circle will be obviously
represented by points of the same region, and every such point can
be considered as a possible position of the "quantum object"
described by our manifold. All manifold's deformations are realized
with equal probabilities (there are no reasons for another
suggestion). Therefore, all possible positions of the point--like
object into the region are realized with equal probabilities. So,
this example shows the possibility of the consideration of above
object as a point with probability description of its positions as
it suggested within standard representation of quantum particles. In
fact, this point is yet a geometrical point only. Below we will
show, how this point becomes the material point. Note also that
parameter $\lambda$ defines the minimal size of a region$\Delta
x\sim \lambda$, where the object can be localized (region at $t=0$).
\par\medskip
\begin{center}{\bf TOPOLOGICAL DEFECT. WAVE CORPUSCULAR PROPERTIES}\end{center}
\par\medskip
Above example does not explain what geometrical properties allow to
differ points of the moving region from neighbour points of the
euclidean space making them observable. To answer at this question
we consider more complex analogy of the closed
4-manifold---two-dimensional torus. In euclidean 3-space such torus
is denoted as topological production of two circles--$S^1\times
S^1$. The role of different manifold's deformations as a reason for
stochastic behaviour was considered in Section 2. Therefore, now we
restrict our consideration to one simplest configuration when one of
$S^1$ is a circle in the plane $XY$ and another is a circle in the
plane $ZX$ (we denote it as $S_1'$).

In pseudueuclidean space this torus looks like a hyperboloid. The
hyperboloid appears if we replace the circle $S_1'$ by a hyperbola
(as it was done in Section 2). Positions of the geometrical object
described by our pseudoeucliden torus are defined by time
cross-sections of the hyperboloid. These positions form an expanding
circles in the two-dimensional euclidean plane (Fig.3).

\begin{center}
\begin{picture}(250,85)
\put(20,45){\vector(1,0){135}}
\put(85,5){\vector(0,1){80}}\thicklines{
\put(85,45){\circle{50}}\put(85,45){\circle{15}}}\thinlines{\put(92,50){$t_1$}
\put(100,62){$t_2$}\put(160,45){x}\put(85,90){y}\put(69,61){\vector(-1,1){15}}\put(101,29){\vector(1,-1){15}}}
\end{picture}
\end{center}
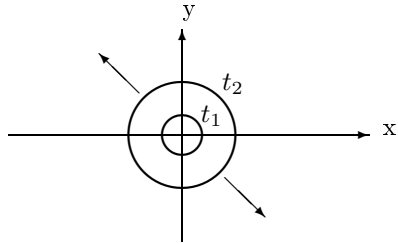
\noindent {\bf FIGURE 3.} Topological defect.
\par\medskip

But we need to have in mind that two-dimensional pseudoeuclidean
torus describes the object existing into two-dimensional space-time
with one-dimensional euclidean "physical" space. This means that an
observable part of the object is represented in our example by the
points of intersections of above circle with $0X$ axis though, as a
whole, the circle is  "embedded" into two-dimensional, "external"
space. This circle can be considered as a topological defect of the
physical one-dimensional euclidean space. Just an affiliation of the
intersection points to the topological object differs these points
geometrically from neighboring points of the one-dimensional
euclidean space. So, in pseudoeuclidean four-dimensional physical
space-time the suggested object described by the Dirac equation
looks like a topological defect of physical euclidean 3-space that
is embedded into 5-dimensional euclidean space, and its intersection
with physical space represents an observable quantum object.

Above analogy with torus does not yet demonstrate appearance of any
wave-corpuscular properties of the object, represented in "physical"
one-dimensional space by the moving intersection point--- properties
that could be expressed by wave function (5) and relation (4). In
the case of considered two-dimensional "space--time" this solution
has the form
$$\psi=u(p)\exp (-2\pi ix^{1}\lambda_{1}^{-1}+2\pi ix^{2}\lambda_{2}).\eqno(10)$$
Topological defect represented by the expanding circle does not
demonstrate any periodical movements when the intersection point
(physical object) propagates along one-dimensional euclidean
$0X$--space.

We will see below that appearance of observable wave-corpuscular
properties is a consequence of nonorientable character of the
topological defect. Torus is a orientable two-dimensional closed
manifold and, therefore, we need to use some nonorientable
low-dimensional analogy. The nonorientable Klein bottle could be
such two--dimensional analogy [6,7]. In the case with torus
topological defect was represented by cross-sections of
pseudoeuclidean torus--plane circles. The Klein bootle is a manifold
that is obtained by gluing of two Mobius strips [10]. Therefore, the
Klein bootle cross-section is an edge of the Mobius strip. This edge
can not be placed in the two--dimensional $XY$--plane without
intersections, and it means that corresponding topological defect is
now a closed curve embedded into three--dimensional $XYZ$--space.

In this case the position of the topological defect relative to its
intersection with $0X$ axis (physical object) can change
periodically. Parameters of this periodical movement depend on
geometrical parameters $\lambda_{1}$, $\lambda_{2}$ (there are no
other parameters with corresponding dimensionality). Such periodical
process can be expressed by the function (10). It leads to the new
interpretation of the wave function as a description of periodical
movement of the topological defect relative to its projection on the
physical space.

In this case corpuscular properties of the above periodical movement
appear as a result of the definition for classical notion of
4-momentum trough the wave characteristic of the topological object,
namely
$$p_{\mu}=2\pi/\lambda_{\mu}.\eqno (11)$$
Substitution of these relations into (18) leads to the Dirac
solution (2)
$$\psi=u(p)\exp (-ip_{1}x^{1}+ip_{2}x^{2}).\eqno(12)$$
It is important to note that within suggested geometrical
interpretation the notions of the less general, macroscopic theory
(4-momentums) are defined by (11) trough the notions of more general
microscopic theory (wave parameters of the defect periodical
movement). This looks more natural than the opposite definitions (4)
within traditional interpretation.
\par\medskip
\begin{center} {\bf GEOMETRIZATION OF ELECTROMAGNETIC WAVES}
\end{center}
\par\medskip

Let us write the Maxwell equations for electromagnetic waves in
vacuum in the symbolic form analogous to the Dirac equation (1).
Namely, we write these equations in the Majorana form [11]
$$i\frac {\partial \textbf{f}^+}{\partial t}=(\textbf{S} \textbf{p}) \textbf{f}^+, \quad \textbf{p} \textbf{f}^+ =0,$$
$$i\frac {\partial \textbf{f}^-}{\partial t}=-(\textbf{S} \textbf{p}) \textbf{f}^-, \quad \textbf{p} \textbf{f}^-
=0,\eqno (13) $$ where
$$\textbf{f}^+=\textbf{E}+i\textbf{H},\quad
\textbf{f}^-=\textbf{E}-i\textbf{H}.\eqno (14)$$ Here $\textbf{E}$
is an electric field, $\textbf{H}$ is a magneyic field,
$\textbf{p}=-i\nabla$ and $\textbf{S}$ is a vector-matrix
$$
S_{x}=\left(\begin{array}{ccc}0&0&0\\0&0&-i\\0&i&0\end{array}\right),
\quad
S_{y}=\left(\begin{array}{ccc}0&0&i\\0&0&0\\-i&0&0\end{array}\right),
\quad
S_{z}=\left(\begin{array}{ccc}0&-i&0\\i&0&0\\0&0&0\end{array}\right).\eqno
(15)
$$

It can be easily seen that Egs.(19) may be rewritten in the symbolic
form analogous to the symbolic form of the Dirac equation (1)
$$i\Gamma^{\mu} \partial_{\mu} \textbf{f}=0,\eqno (16)$$ where
bivector \textbf{f} and matrix $\Gamma^{\mu}$ have the form
$$
\Gamma^0=\left(\begin{array}{cc}0&1\\1&0\end{array}\right), \quad
\Gamma^{1,2,3}= \left(\begin{array}{cc}0&-{\bf S }\\{\bf
S}&0\end{array} \right),\quad
\textbf{f}=\left(\begin{array}{cc}\textbf{f}^+\\\textbf{f}^-\end{array}\right).
$$ We write here six-row matrices trough three-row ones). We see
that Egs.(16) has formally the same form as the Dirac equation (1)
with $m=0$. Only instead of the Dirac bisinor $\psi$ we have here
bivector \textbf{f}, and instead Dirac matrices $\gamma^{\mu}$ we
have matrices $\Gamma^{\mu}$. As Maxwell's Egs.(16) looks formally
like Dirac's Egs.(1) it seems reasonable to use for their
geometrization the same arguments as we used for the Dirac equation
geometrization.

For plane waves the solution of Eq.(16) has the form
$$\textbf{f}^+=\textbf{f}^+_k\exp i(\textbf{kr}-\omega t),\quad \textbf{f}^-=\textbf{f}^-_k\exp i(\textbf{kr}-\omega t),\quad \omega = |\textbf{k}|.
\eqno (17)$$ As for topological interpretation of solution (2) of
the Dirac equation we suggest that function (17) does not define any
wave process into the space but describes periodical movement of the
space topological defect. As function (2) solution (17) can be also
considered as the realization of fundamental group of some closed
topological manifold. Due to exponential factor in (17) this group
contains the translation group (as for solution (2)), but now our
solution are complex bivectors--not a bispinors. These vectors does
not realize the representation of reflections along three different
axes as it was for bispinors. These vectors $\textbf{f}^+$ and
$\textbf{f}^-$ consist of axial and polar vectors (see (14)) and
thus these vectors are transformed one into another only in result
of the reflection of space axes. Therefore, solution (17) realizes
representation of a sliding symmetry in 4-space only along
time--axis. This distinguishes the supposed fundamental group from
the fundamental group considered in previous Sections (four sliding
symmetries along four perpendicular axes).

There is no investigation in topology, where 4-manifolds with above
fundamental group were considered. So, we again can establish
connections between geometrical properties of the manifold and
observable physical properties of electromagnetic waves only using
low-dimensional analogies. Wave-corpuscular dualism of
electromagnetic waves and possibility of stochastic behavior can be
demonstrated in the same manner as in Sections 2,3. But
electromagnetic waves have some additional important
property---their velocity does not depend on the source motion. We
will show below how geometrical properties of the closed topological
4-manifold can explain this fact.

Suppose that 4-manifold corresponding to electromagnetic wave has
the form $M^3(\textbf{r})\times M^1(\textbf{r},t)$, that is it can
be represented as a product of nonorientable three-dimensional
euclidean closed manifold $M^3(\textbf{r})$ and one-dimensional
manifold $M^1(\textbf{r},t)$ homeomorphic to a pseudoeuclidean
circle. The formal reason for such representation is the
distinguished role of the euclidean space within fundamental group:
only into euclidean subspace translation group is combined with
reflections. Consider now a low-dimensional analogy that explains an
independence of light velocity on the source motion.

Instead of four-dimensional manifold $M^3(\textbf{r})\times
M^1(\textbf{r},t)$ we consider, as in previous Section,
two-dimensional analogy---manifold $S^1\times S_1'$, where $S^1$ is
a one-dimensional euclidean circle and $S_1'$ is a pseudoeuclidean
circle. This manifold was considered in Section 3, and it looks like
a hyperboloid. For electromagnetic waves $m=0,E=cp$. Within our
notation it leads to relation $p_{1}=p_{2}=p$ or
$\lambda_{1}=\lambda_{2}=\lambda$. Therefore, there have to be only
one parameter with dimensionality of length, and this will be the
case if $S_t^1$ is a pseudoeuclidean circle of zeroth radius.
Equation for such circle has the form $x^{2}-t^{2}=0$, and
hyperboloid is transformed into a cone (Fig.3). This means that the
points representing in this example electromagnetic wave move with
velocity equals $\pm1$ ($\pm c$--in chosen units system), and this
velocity does not depend on coordinate frame rotations (does not
depend on transfer from one moving inertial frame to another). From
topological point of view this result is a consequence of the fact
that the zeroth radius can be considered as topological invariant.
Therefore, within geometrical approach light velocity appears to be
topological invariant of the manifold representing electromagnetic
wave, and this is the reason of its independence of the source
motion.
\par\medskip
\begin{center}{\bf NEW OPPORTUNITIES}\end{center}
\par\medskip

Let us notice shortly some new opportunities that arise within
suggested topological concept. For example, including into physical
world the objects from outer five-dimensional space can explain
rather simply the origin of nonlocal correlation in
EPR--experiments. We see at Fig.3 that two noninteracting identical
particles (intersection points) can move in opposite directions
being parts of the same topological defect (the circle). In the act
of measurement of one of the particles the symmetry of the whole
topological defect can be changed (for example, it can become
nonorientable manifold being at first orientable  one). And another
particle, being the part of the same defect, would "feel" at once
this change without any interaction (duration of above symmetry
change is defined by measurement procedure, and it does not depend
on the distance between particles).

The more important consequence of new approach is the possibility to
overcome the difficulties of physics of many-electron atoms
connected with the "many-body" problem.  Earlier we suggested
geometrical interpretation of the Dirac equation for hydrogen atom
[1]. The electron Coulomb potential appears to be within this
interpretation the connectivity of the universal covering space of
corresponding topological defect and has nothing to do with any
negative particle inside the atom. It means, may be, that there is
no necessity to consider atoms as many-particle systems because
many--particles potentials have to be changed within geometrical
approach by corresponding connectivities, and these coonectivities
are functions of only one space-time coordinate. Then new possible
relativistic equations (instead of Schrodinger ones) have to be
"one-particle" equations for functions of one space-time coordinate.
We are now looking for such new possible potentials and equations.
\par\medskip
\begin{center} {\bf REFERENCES}\end{center}

\noindent 1. O.A. Olkhov, {\it Journ.of Phys.:Conf.Ser}.,{\bf
67},012037(2007); arXiv: 0706.3461.\\
2. J.D. Bjorken, S.D. Drell, {\it Relativistic Quantum Mechanics
and Relativistic Quantum Fields}, New York McGraw-Hill, 1964.\\
3. A.I. Achiezer, S.V. Peletminski, {\it Fields and Fundamental
Ineractions}, Kiev Naukova Dumka, 1986, Ch.1.\\
4. P. L. Rachevski, {\it Riemannian geometry and tensor analysis}, Мoscow, Nauka, 1966, \S 55.\\
5. V.A. Jelnorovitch, {\it Theory of spinors and its applications}, Мoscow, August-Print, 2001, \S 1.3.\\
6. M. Hamermesh, {\it Group Theory and Its Application to Physical
Problems}, Argonne National Laboratory, 1964. \\
7. H.S.M. Coxeter, {\it Introduction to Geometry}, N.Y.-London John
Wiley and Sons, 1961. \\
8. B.A. Dubrovin, S.P. Novikov, A.T. Fomenko, {\it Modern geometry},
Moscow, Nauka, 1986. \\
9. A.S. Schvartz {\it Quantum Field Theory and Topology}, Moscow,
Nauka, 1989.\\
10. D. Gilbert, S. Kon-Fossen, {\it Nagladnaya geometriya (rus) },
Moscow, Nauka, 1981.\\
11. А.I. Achiezer, V.B. Berestetski, {\it Quantum Electrodynamics},
Moscow, Nauka, 1981.\\

\end{document}